\documentclass[conference]{IEEEtran}
\usepackage{cite}
\usepackage{amsmath,amssymb,amsfonts}
\usepackage{algorithmic}
\usepackage{graphicx}
\usepackage{textcomp}
\usepackage{xcolor}
\usepackage[breaklinks, hidelinks]{hyperref}
\usepackage[nolist, nohyperlinks]{acronym}
\usepackage{orcidlink}
\usepackage{adjustbox}
\usepackage{tikz}
\usepackage{colortbl}
\usepackage[caption=false]{subfig}

\begin{acronym}
\acro{AE}{autoencoder}
\acro{ASV}{automatic speaker verification}
\acro{BCE}{binary cross-entropy}
\acro{CM}{countermeasure}
\acro{CNN}{convolutional neural network}
\acro{COTS}{commercial-off-the-shelf}
\acro{CQCC}{constant-Q cepstral coefficient}
\acro{CRNN}{convolutional recurrent neural network}
\acro{DNN}{deep neural network}
\acro{DF}{deepfake}
\acro{EER}{equal error rate}
\acro{ELU}{exponential linear unit}
\acro{FCN}{fully convolutional network}
\acro{FFT}{fast Fourier transform}
\acro{GMM}{Gaussian mixture model}
\acro{GRU}{gated recurrent unit}
\acro{ID}{identifier}
\acro{IoT}{Internet of things}
\acro{LA}{logical access}
\acro{LSTM}{long short-term memory}
\acro{MVDR}{minimum variance distortionless response}
\acro{MWF}{multichannel Wiener filtering}
\acro{ReMASC}{Realistic Replay Attack Microphone Array Speech}
\acro{RNN}{recurrent neural network}
\acro{ROC}{receiving operating characteristic}
\acro{TTS}{text-to-speech}
\acro{PA}{Physical access}
\acro{PFA}{probability of false alarm}
\acro{SOTA}{state-of-the-art}
\acro{STFT}{short-time Fourier transform}
\acro{VA}{voice assistant}
\acro{VC}{voice conversion}
\end{acronym}

\definecolor{Gray}{gray}{0.85}
\definecolor{red_cool}{rgb}{0.5, 0.0, 0.0}

\def\BibTeX{{\rm B\kern-.05em{\sc i\kern-.025em b}\kern-.08em
    T\kern-.1667em\lower.7ex\hbox{E}\kern-.125emX}}
\begin{document}

\title{Impact of Microphone Array Mismatches to Learning-based Replay Speech Detection
\thanks{}
}

\author{\IEEEauthorblockN{Michael~Neri~\orcidlink{0000-0002-6212-9139}, Tuomas~Virtanen~\orcidlink{0000-0002-4604-9729}}
\IEEEauthorblockA{\textit{Faculty of Information Technology and Commmunication Sciences}, \\ \textit{Tampere University}, Tampere, Finland}
\IEEEauthorblockA{\{michael.neri, tuomas.virtanen\}@tuni.fi}
}

\maketitle

\begin{abstract}
In this work, we investigate the generalization of a multi-channel learning-based replay speech detector, which employs adaptive beamforming and detection, across different microphone arrays. In general, deep neural network-based microphone array processing techniques generalize poorly to unseen array types, i.e., showing a significant training-test mismatch of performance. We employ the ReMASC dataset to analyze performance degradation due to inter- and intra-device mismatches, assessing both single- and multi-channel configurations. Furthermore, we explore fine-tuning to mitigate the performance loss when transitioning to unseen microphone arrays. Our findings reveal that array mismatches significantly decrease detection accuracy, with intra-device generalization being more robust than inter-device. However, fine-tuning with as little as ten minutes of target data can effectively recover performance, providing insights for practical deployment of replay detection systems in heterogeneous automatic speaker verification environments.
\end{abstract}


\section{Introduction}
Recently, \acp{VA} have been employed for human-machine interaction, exploiting voice as a biometric trait for authorizing the user~\cite{Huang_IoTJ_2022}. In fact, using \acp{VA} we can control smart devices in our homes through \ac{IoT} networks and transmit sensitive information over the Internet. Malicious users can employ attacks on audio recordings~\cite{Liu_TASLP_2023} to access personal data and devices by deceiving the \ac{ASV} of \acp{VA}. By applying \ac{TTS} and/or \ac{VC} algorithms, it is possible to perform a \ac{LA} attack which modifies both talker traits or speech content. In the same direction, compressing or quantizing speech recordings can mask \ac{LA} attacks by introducing artifacts, namely \ac{DF} attack~\cite{Liu_TASLP_2023}. \ac{PA} attacks focus on deceiving the \ac{ASV} system at the microphone level~\cite{Kinnunen_ICASSP_2017}. An attacker can mimic user's speech to access to sensitive data or use a spoofing microphone to later replay the recorded speech to the \ac{ASV}, which are called \textit{impersonation attack}~\cite{Huang_IoTJ_2022} and \textit{replay attacks}~\cite{Gong_SPL_2020} respectively. We focus on \textit{replay attack} as, differently from other biometric markers, voice is easy to be captured by attackers to from the target speaker~\cite{Delac_ISEM_2004}. Moreover, \ac{ASV} struggles to distinguish between genuine and replay speech even with \ac{COTS} devices like smartphones and loudspeakers~\cite{Neri_2025_ARXIV, Gong_Interspeech_2019}.

\begin{figure*}[ht!]
    \centering
    \includegraphics[width=0.8\linewidth]{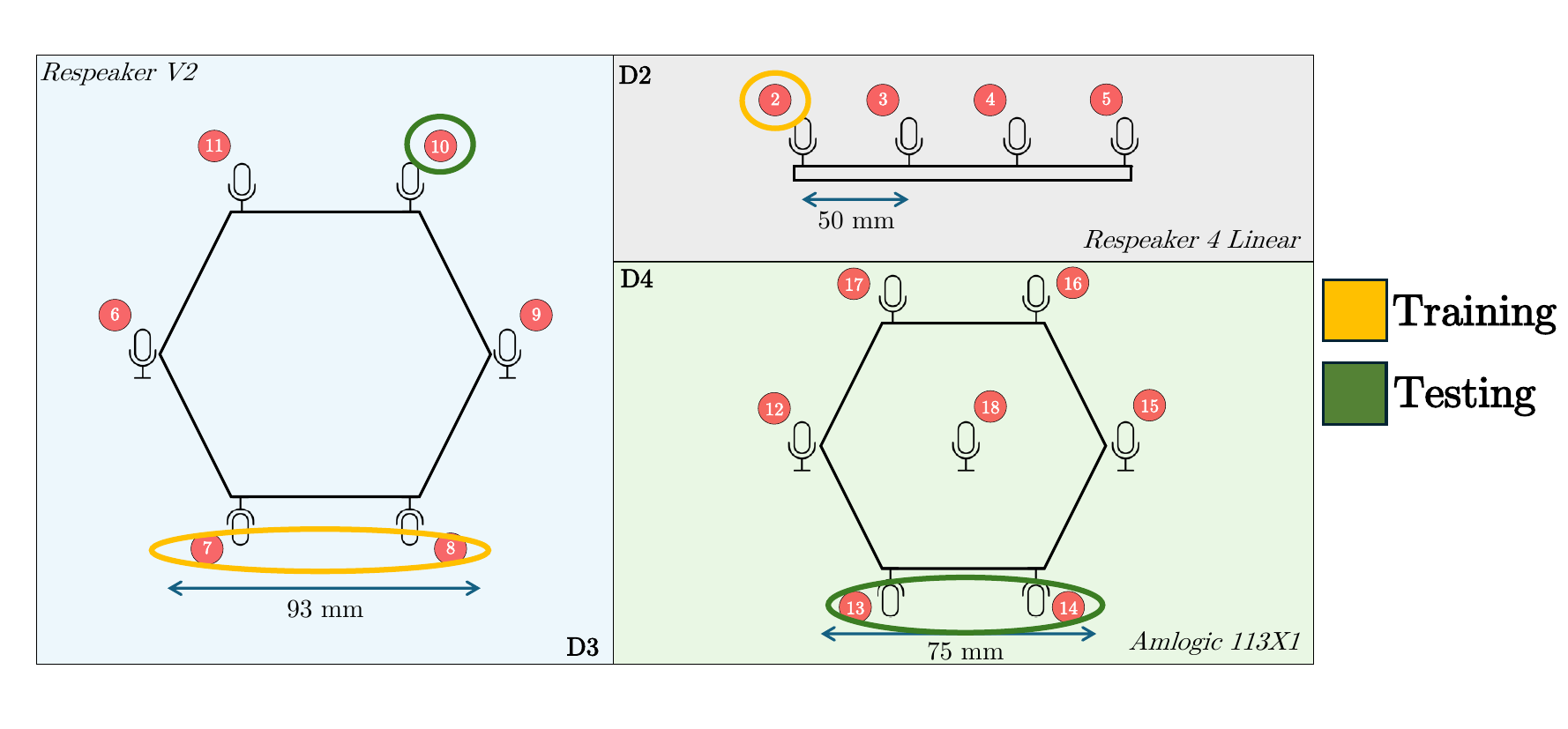}
    \caption{Overview of microphone arrays from ReMASC~\cite{Gong_Interspeech_2019} dataset involved in the experiments. Few examples of training and testing configurations are depicted. For instance, a model is trained on microphone with \ac{ID} $2$ from $\mathrm{D}2$ and tested on microphone \ac{ID} $10$ from $\mathrm{D}3$ with speech recordings in the same environment. An example of multi-channel setup is training a model on the couple $(7, 8)$ from $\mathrm{D3}$ and testing on the couple $(13, 14)$ from $\mathrm{D}4$.} 
    \label{fig:Mics_Desc}
\end{figure*}

Prior works aimed to tackle \ac{PA} attacks by collecting single-channel recordings, such as RedDots~\cite{Kinnunen_ICASSP_2017}, ASVSpoof2017 \ac{PA}~\cite{kinnunen2017asvspoof}, ASVSpoof2019~\cite{todisco2019asvspoof} \ac{PA}, and ASVSpoof2021 \ac{PA}~\cite{Liu_TASLP_2023}. From these datasets, several single-channel replay attack detection models have been designed, employing both \ac{DNN}~\cite{Luo_ICASSP_2021, Xue_LSP_2024} and hand-crafted time-frequency features~\cite{boyd2023voice, Xu_TASLP_2023}. However, single-channel replay detectors demonstrated poor generalization capabilities across different environments and devices, similarly to what happens in different tasks like in acoustic scene classification and unsupervised anomalous sound detection~\cite{heittola2020acoustic, Dohi2022}. To cope with this issue, previous works collected more diverse datasets~\cite{heittola2020acoustic, Dohi2022}, encompassing different environments and microphones, or defining ad-hoc training procedures to minimize the domain shift~\cite{Neri_2024_ICMEW}.

For speech enhancement and separation tasks, microphone arrays are used in \ac{ASV} systems to exploit spatial information and improve audio quality~\cite{omologo2001speech}. However, training-testing mismatches of microphone arrays is typically a severe issue in \ac{DNN}-based microphone array processing, thus yielding poor generalization capabilities to unseen testing devices. To study the impact of these training-testing differences, in~\cite{vincent2017analysis} the authors conducted initial studies on mismatched noise environments, different microphones, simulated, and real data for speech enhancement and separation. In~\cite{Lin_ICASSP_2024} a fixed beamformer in conjuction with a \ac{RNN} has been analyzed in terms of speech enhancement performance changing the arrangement of microphones in an array. Similarly, Han \textit{et al.}~\cite{Han_ICASSP_2024}  designed an unsupervised method for adapting a single-channel \ac{DNN} to seen array configurations. In the context of speaker diarization, a \ac{DNN} has been designed to be robust to array mismatch in~\cite{Mariotte_TASLP_2024}. Recent works have explored to include array geometry information to \ac{DNN} to improve robustness of learning-based approaches to unseen devices~\cite{Heikkinen_ICASSP_2024, Pan_TASLP_2025}.

Regarding the replay speech detection attack, the lack of datasets and models on this task  makes the microphone array mismatch more challenging also with unseen factors such as environments~\cite{Neri_2025_ARXIV} and different array geometries. To the best of our knowledge, studying the impact of microphone mismatches with diverse devices and number of microphones has never been explored in the \ac{ASV} context. 

To better study this phenomenon, we reply to the following research questions:
\begin{itemize}
    \item \textit{RQ1: How much is the degradation of the detection performance of a learning-based detector with microphone array mismatch?}
    \item \textit{RQ2: Is fine-tuning a learning-based detector trained on device $\mathrm{D}_i$ to target device $\mathrm{D}_j$ beneficial?}
    \item \textit{RQ3: How much data is required to recover the lower-bound \ac{EER} of the target array from a pre-trained model on a source array (assuming that the lower-bound is a model trained solely on target data)?}
\end{itemize}

To answer these research questions, we employ \ac{ReMASC} dataset~\cite{Gong_Interspeech_2019}, which is the only replay speech dataset that encompasses real recordings from different microphone arrays. Moreover, we employ the state-of-the-art replay detector M-ALRAD~\cite{Neri_2025_ARXIV}, which combines adaptive beamforming and detection, achieving the best detection performance. 

The structure of the paper is organized as follows. Section~\ref{sec:materials} depicts the model and the dataset used for evaluation of microphone array mismatch for replay speech detection. Section~\ref{sec:results} shows the results of the performed analysis, changing the number of microphones involved in an array and varying the training and the testing setup. Finally, Section~\ref{sec:conc} draws the conclusions and future directions to mitigate the problem.

\section{Materials}\label{sec:materials}
This section describes the \ac{ReMASC} dataset and the model M-ALRAD (Multi-channel Replay Attack Detector) employed in the evaluation of performance in mismatched microphone arrays.

\subsection{Dataset under analysis}
\ac{ReMASC}~\cite{Gong_Interspeech_2019} has been selected since it is the only dataset providing multi-channel recordings, which have been collected by devices with different number of microphones and geometries, for this task. All recordings are synchronized across four parallel \ac{ASV} microphone arrays ${\mathrm{D}1, \mathrm{D}2,\mathrm{D}3,\mathrm{D}4}$ with $2$, $4$, $6$, and $7$ omnidirectional microphones, respectively, with $9240$ genuine and $45472$ replay audio samples. Moreover, $\mathrm{D}4$'s sampling rate is $16$ kHz, differently from $\mathrm{D}1$, $\mathrm{D}2$, and $\mathrm{D}3$ which operate at $44.1$ kHz. Bit depth is $16$ bits for all devices, except for $\mathrm{D}3$ which is $32$ bits. Overall, \ac{ReMASC} features $55$ diverse speakers, varying in gender and vocal characteristics, and includes recordings from four distinct environments: an outdoor setting (Env-A), two enclosed spaces (Env-B and Env-C), and a moving vehicle (Env-D), capturing different acoustic conditions. The dataset employs two spoofing microphones to record genuine speech, which is then replayed through four playback devices of varying quality. Figure~\ref{fig:Mics_Desc} illustrates the structure of microphone arrays and the identification code for each single-channel omnidirectional microphone. To select a microphone from the dataset, we label each microphone with an \ac{ID} from $2$ to $18$.

\subsection{Selected replay speech detector}
In our experiments, we employ M-ALRAD~\cite{Neri_2025_ARXIV}, which is a \ac{CRNN}-based approach that jointly processes $N$ multi-channel complex \acp{STFT} $\{X_{{\mathrm{STFT}}_{n,T,F}}, n = 1, \ldots, N\}$ with $T$ and $F$ time and frequency bins, respectively, to produce a single-channel beamformed spectrogram for detecting replay speeches. First, a \ac{CNN} $f_{BM}:\mathbb{C}^{N \times T \times F} \rightarrow \mathbb{C}^{T \times F}$ outputs beamforming weights $\mathbf{W} \in \mathbb{C}^{N \times T \times F}$ through a series of Conv2D-BatchNorm-ELU-Conv2D operations, where real and imaginary parts are concatenated along the channel dimension. The beamformed spectrogram $\hat{X}_{\mathrm{STFT}}$ is computed as

\begin{equation} \hat{X}_{{\mathrm{STFT}_{T, F}}} = \sum_{n} X_{{\mathrm{STFT}}_{n,T,F}} \cdot \mathbf{W}_{n,T,F}. \end{equation}
The beamformed spectrogram is then processed by a \ac{CRNN} classifier, previously used for speaker distance estimation~\cite{Neri_TASLP_2024, Neri_WASPAA_2023}. Magnitude and phase features of the beamformed \ac{STFT} are extracted, with sin\&cos transformations applied to the phase. These features are arranged in a $T \times F \times 3$ tensor and passed through three convolutional layers with $1 \times 3$ filters and batch normalization, followed by parallel max and average pooling. Two bi-directional \ac{GRU} layers with $128$ neurons refine the feature maps, and the final hidden state is mapped to the binary prediction $\hat{y} \in \mathbb{R}$ exploiting a fully connected layer with softmax activation function, where $0$ denotes the genuine class and $1$ represents the replay class.

The model is trained to minimize the binary cross-entropy loss between predicted and true binary classes, i.e., genuine or replay. In addition, orthogonality and sparsity losses are employed using the same hyperparameters as in~\cite{Neri_2025_ARXIV}.

\section{Experimental results}\label{sec:results}
In this section, we evaluate the detection performance of M-ALRAD when different microphone configurations are used in the training and testing sets. First, details on the analysis regarding the subset of \ac{ReMASC} and training hyperparameters are provided. Then, we first evaluate a single-channel mismatch scenario. Next, we increase the number of microphones from two to four, in order to reply to RQ1. Finally, we perform fine-tuning to provide some insights to RQ2 and RQ3.

\subsection{Metric and implementation details}
We employed recordings from Env-B where the configuration of talker/loudspeaker and microphone are varied in a grid whereas the room is the same. The design rationale of this choice is to focus on microphone array mismatches without the influence of varying the environment. The selected microphone arrays from \ac{ReMASC} are $\{\mathrm{D}2,\mathrm{D}3,\mathrm{D}4\}$ since $\mathrm{D}1$ does not have genuine speech recordings due to hardware fault during data collection~\cite{Gong_Interspeech_2019}. 

Following~\cite{Neri_2025_ARXIV}, each model is trained or fine-tuned for each setup of microphones with a batch size of $32$ using a cosine annealing scheduled learning rate of $0.001$ for $50$ epochs. We only analyze the first second of multi-channel recordings to reduce the computational complexity of the approach~\cite{Gong_SPL_2020, Neri_2025_ARXIV}. The \ac{STFT} is computed with a Hanning window of length $32$ and $46$ ms with $50 \%$ overlap for $44.1$ and $16$ kHz, respectively. When sampling rate are different, e.g., training on $\mathrm{D2}$ and testing on $\mathrm{D4}$ and viceversa, upsampling or downsampling is performed on the target recording in order to use the same \ac{STFT} parameters. To evaluate the replay speech attack detection, the \ac{EER} metric is used following the same train-test split provided by the authors of the ReMASC dataset~\cite{Gong_Interspeech_2019}. Five independent runs are collected to compute the $95\%$ mean confidence intervals. Overlapping training and testing channels are not performed, e.g., training on $(2, 3)$ and testing on $(3, 4)$, for avoiding data leakage.

\subsection{Single-channel mismatch}
\begin{figure}[ht!]
    \centering
    \includegraphics[width=1\linewidth]{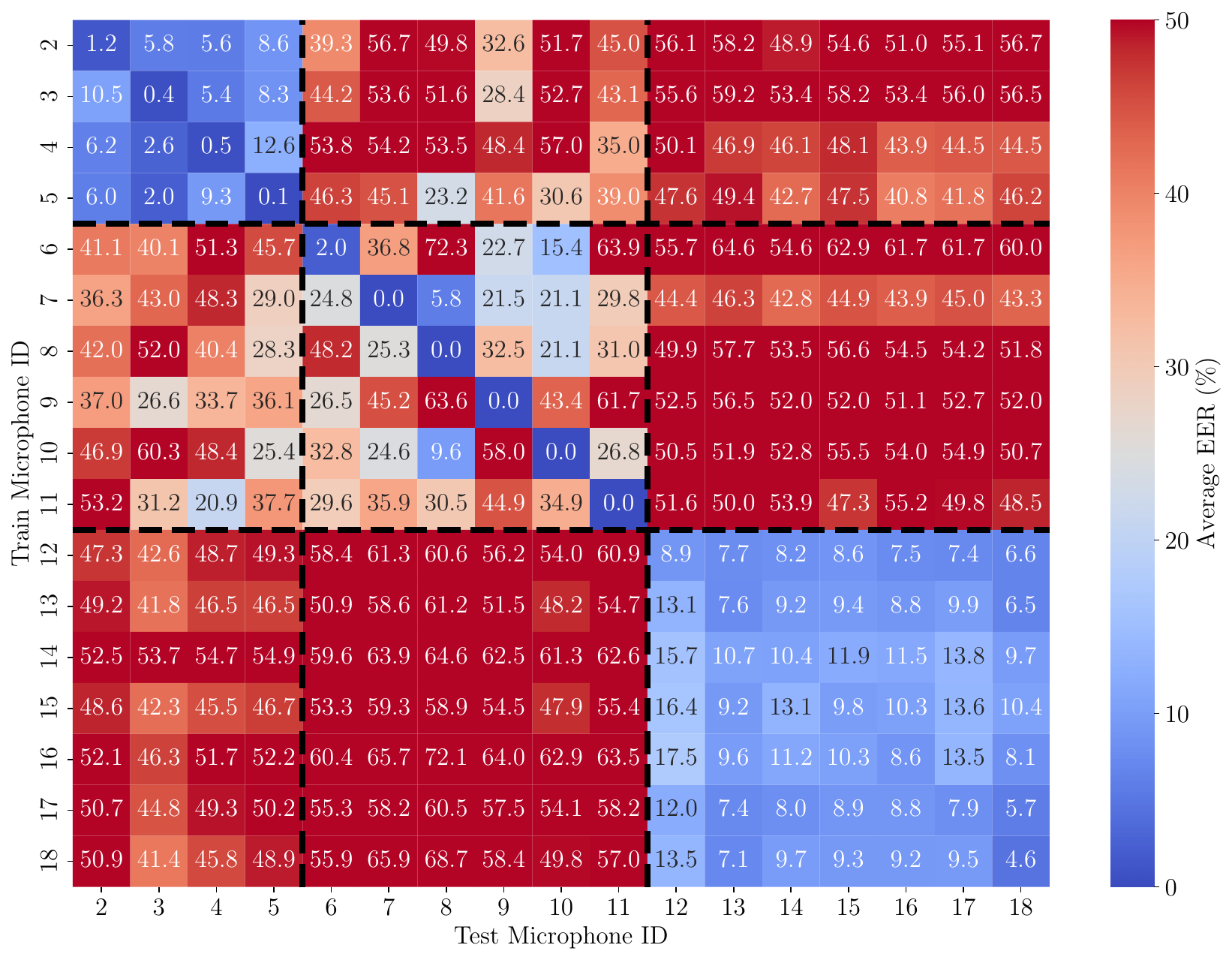}
    \caption{Single-channel cross training-testing \acp{EER}.}
    \label{fig:CrossTrainingSingleChannel}
\end{figure}

\begin{figure}[ht!]
    \centering
    \includegraphics[width=0.9\linewidth]{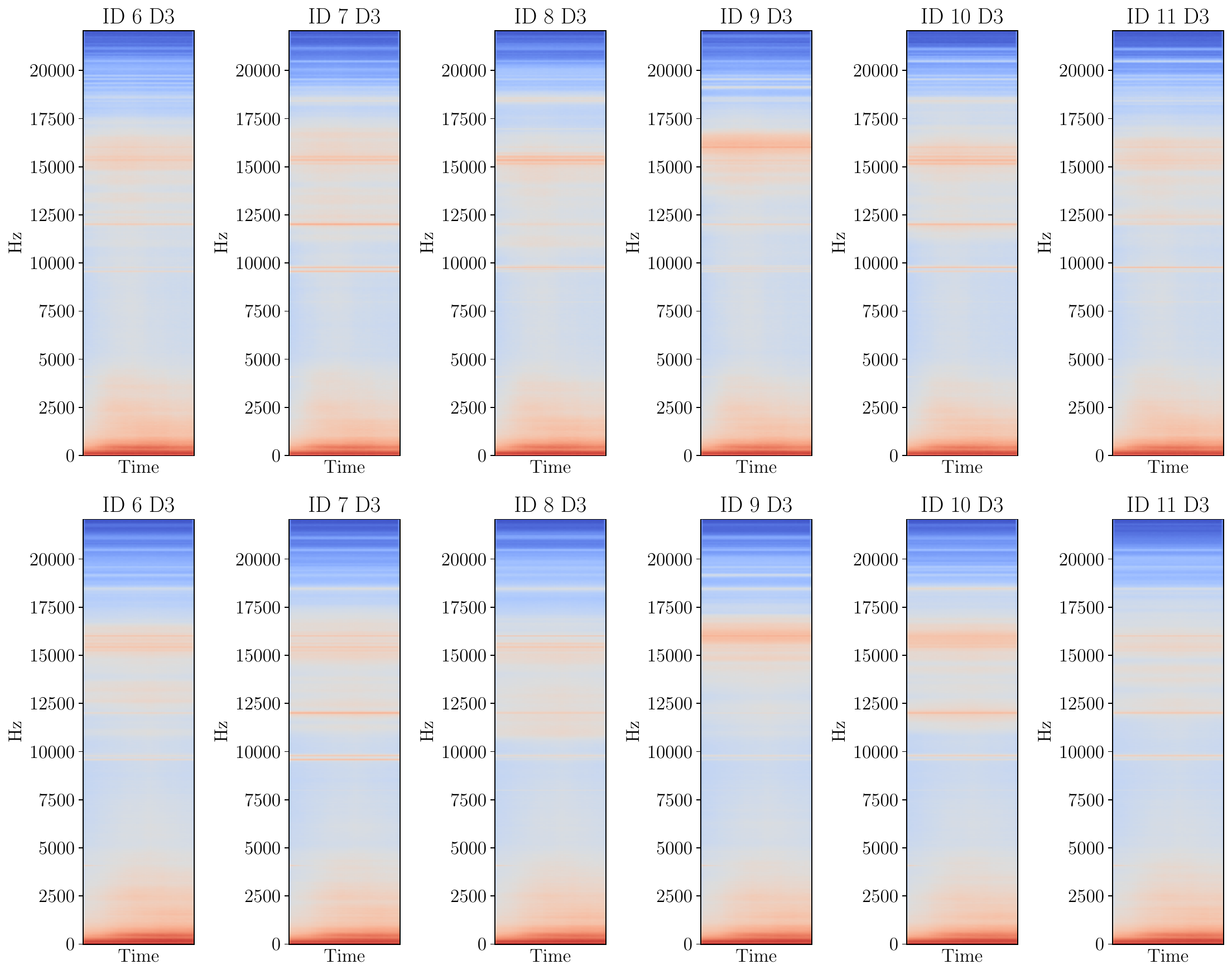}
    \caption{Average \ac{STFT} magnitudes from $\mathrm{D}3$ in Env-B. The first row consists of genuine speeches whereas the second row consists of replay speeches.}
    \label{fig:AverageSpectraD3}
\end{figure}

First, we evaluate the performance of training a model on a single microphone and testing on a different one. By doing so, it is possible to jointly assess (i) \textit{inter-device} mismatches, e.g., training on microphone \ac{ID} $2$ and testing on microphone \ac{ID} 4 of array $\mathrm{D}2$, and (ii) \textit{intra-device} mismatches, e.g., training on microphone with \ac{ID} $1$ from array $\mathrm{D}2$ and testing on microphone with \ac{ID} 6 from array $\mathrm{D}3$. All the results have been collected in a matrix visualization in Figure~\ref{fig:CrossTrainingSingleChannel}, where rows denote the training microphone \ac{ID} and columns are the testing microphone \ac{ID}. It is clear from the results that \textit{inter-device} experiments show better performance, as microphones belonging to the same array share similar acoustic characteristics like frequency responses, as previously demonstrated in prior works~\cite{vincent2017analysis}. Instead, \textit{intra-device} results are worse than random guess (i.e., \ac{EER} $ \geq 50\%$). In addition, microphones belonging to $\mathrm{D}3$ on average performs worse than $\mathrm{D}2$ and $\mathrm{D4}$ both in the \textit{intra-} and \textit{inter-device} scenarios. To analyze this behavior, the average \ac{STFT} magnitude of each microphone in array $\mathrm{D}3$ is depicted in Figure~\ref{fig:AverageSpectraD3}, which shows significant differences between microphones, depicting dominant spectral peaks which are different for different microphones. As the model is trained using features derive from the \ac{STFT}, these difference cause domain shift in \textit{intra-device} results.

\subsection{Multi-channel mismatch}

\begin{figure}[ht!]
    \centering
    \includegraphics[width=1\linewidth]{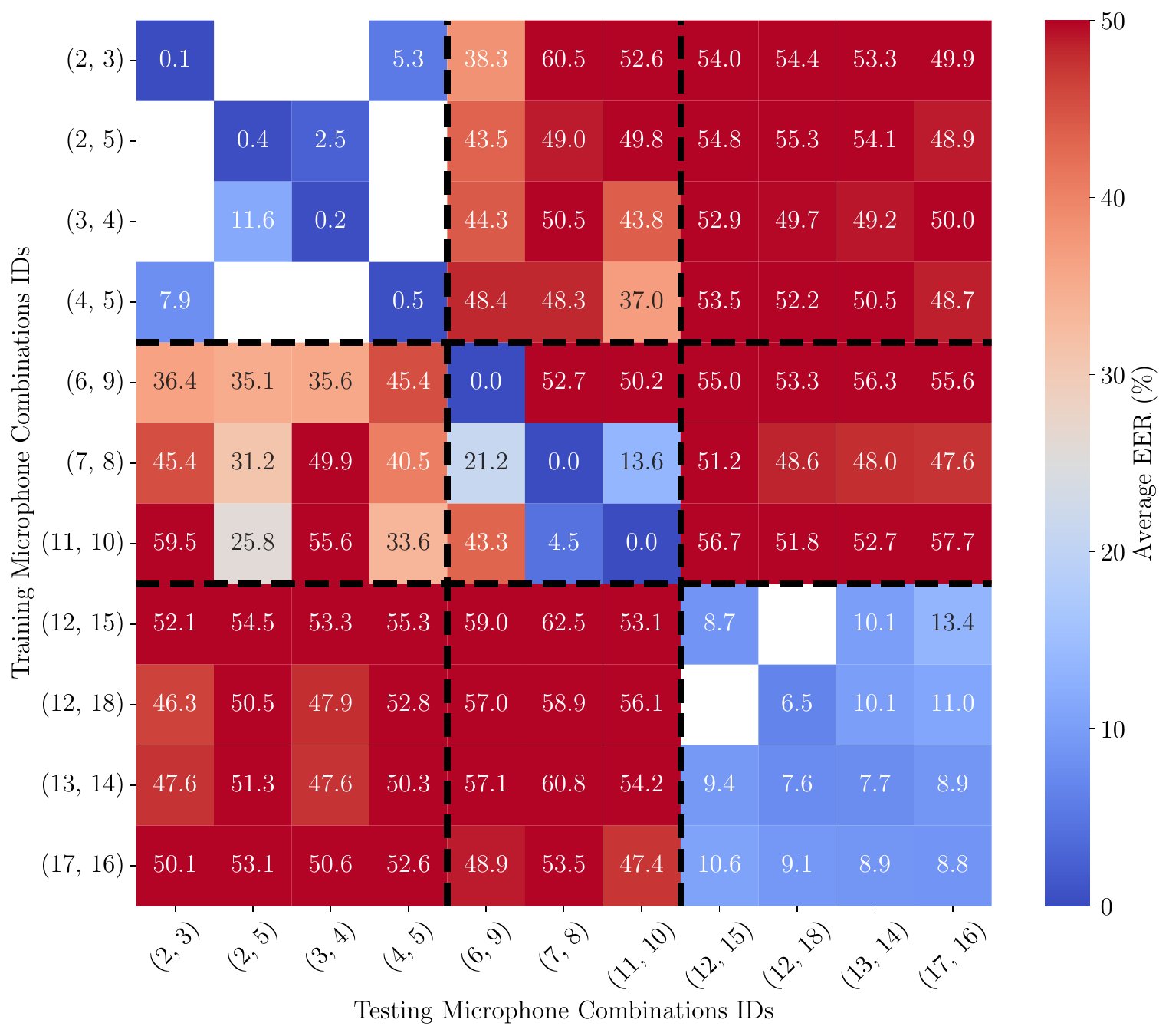}
    \caption{Two channels cross training-testing \acp{EER}.}
    \label{fig:CrossTrainingTwoChannel}
\end{figure}

With two-channel setups in Figure~\ref{fig:CrossTrainingTwoChannel}, the overall error rates in matched conditions decrease compared to single channel, indicating that combining two microphone inputs helps in improving the detection performance. Logically, \textit{intra-device} pairs continue to show the best results. When switching to significantly different pairs, such as $(6,9)$ to $(12,15)$, the degradation is more evident, reinforcing that spatial and array-level shifts pose a difficult generalization challenge.

\begin{figure}[ht!]
    \centering
    \includegraphics[width=0.9\linewidth]{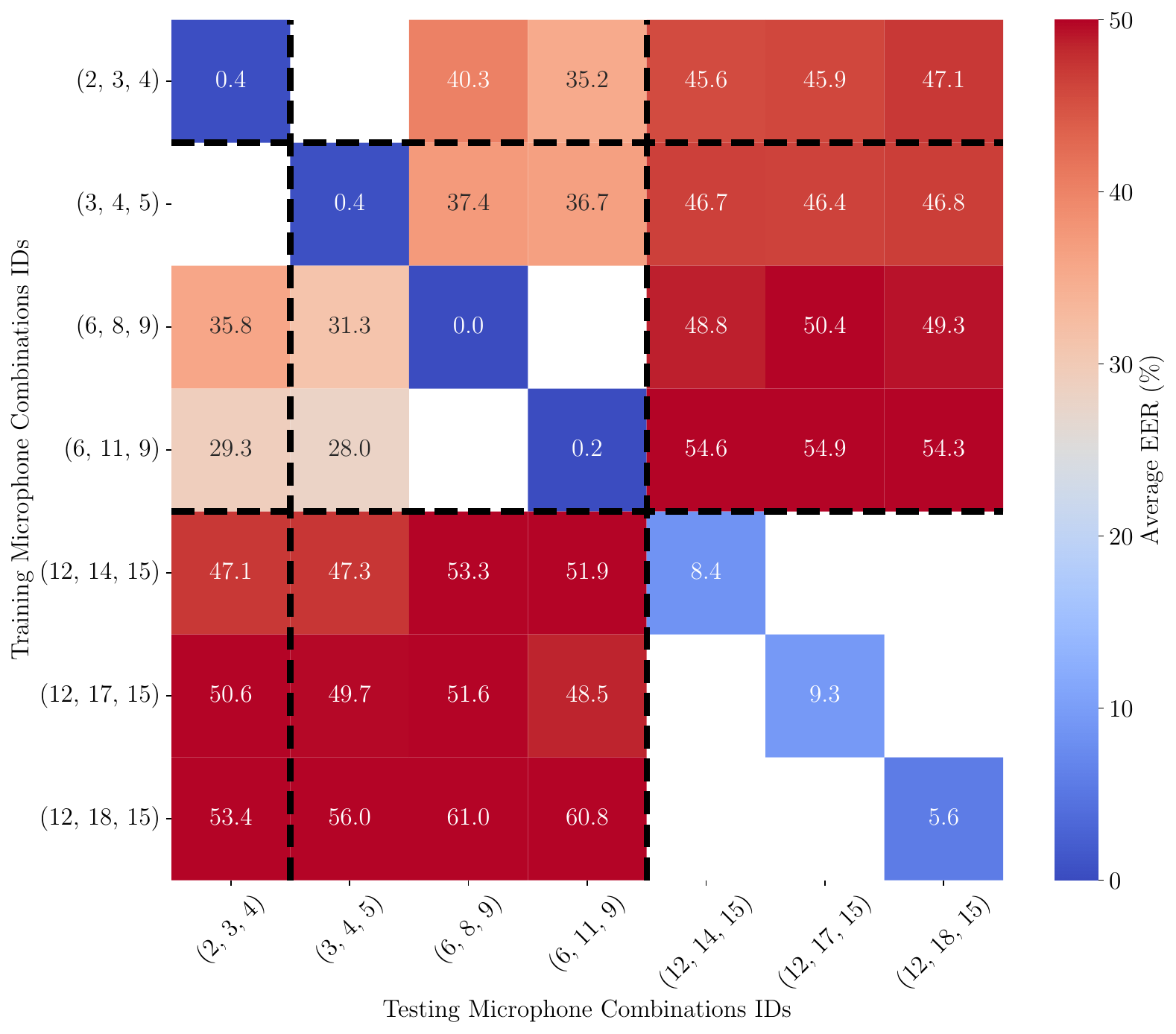}
    \caption{Three channels cross training-testing \acp{EER}.}
    \label{fig:CrossTrainingThreeChannel}
\end{figure}

The three-channel matrix in Figure~\ref{fig:CrossTrainingThreeChannel} shows a similar trend with respect to the two channel configuration. In fact, for instance training on $(2,3,4)$ from $\mathrm{D}2$ and testing on $(12,17,15)$ from $\mathrm{D}4$, still result in performance drops and random guess predictions.

\begin{figure}[ht!]
    \centering
    \includegraphics[width=1\linewidth]{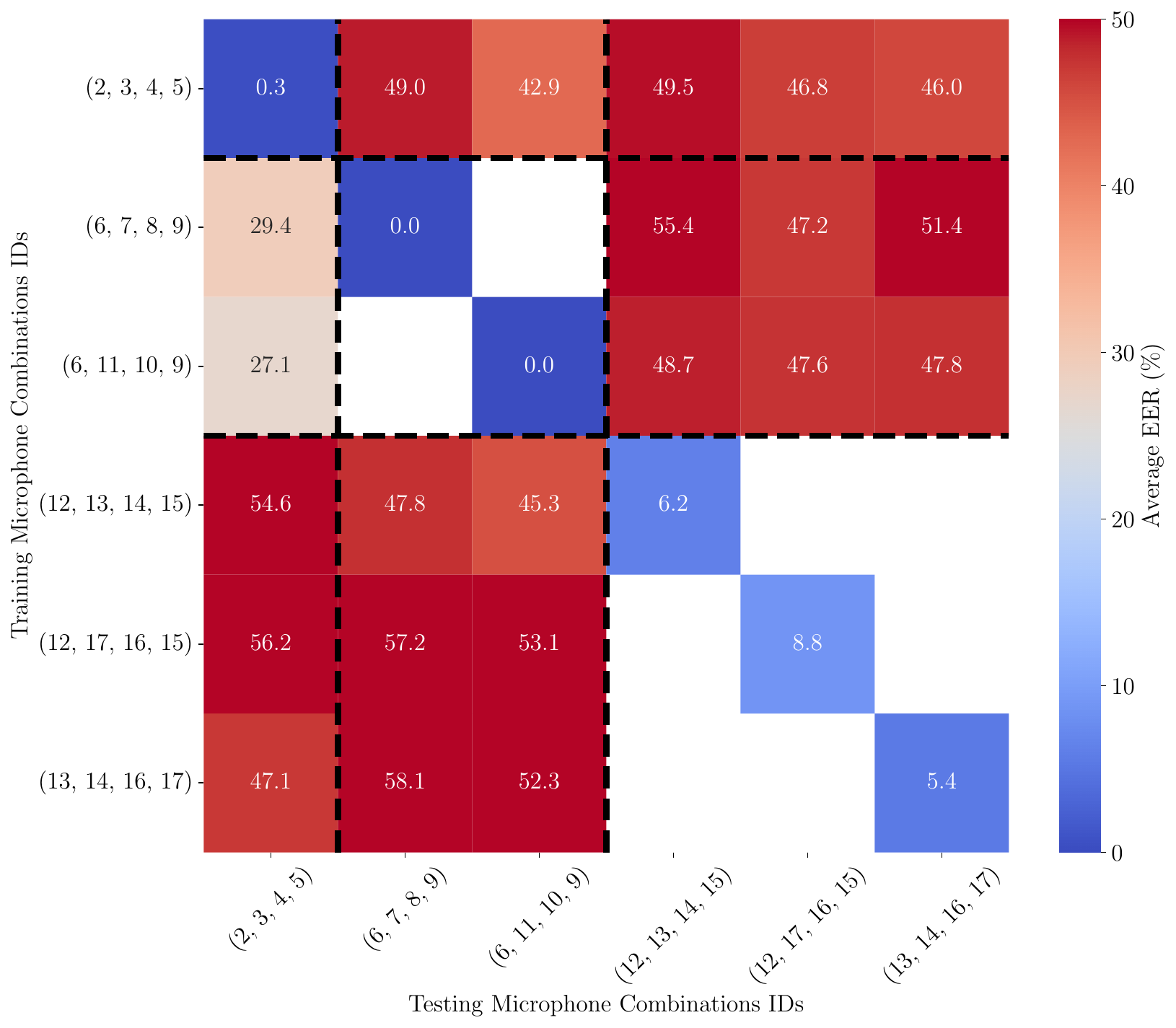}
    \caption{Four channels cross training-testing \acp{EER}.}
    \label{fig:CrossTrainingFourChannel}
\end{figure}

The four-channel matrix in Figure~\ref{fig:CrossTrainingFourChannel} reveals that \textit{intra-device} configurations continue to achieve the lowest error rates. However, non-random guess performance are achieved by training M-ALRAD on four microphones from $\mathrm{D}2$ and testing on $\mathrm{D}3$. Clearly, generalizing on the array $\mathrm{D4}$ is more challenging since this device operates at a different sample rate.

\begin{figure*}[ht!]
    % First Image
    \centering
    \subfloat[Source \ac{ID} $2$ and target \ac{ID} $12$.]{\includegraphics[width=0.25\linewidth]{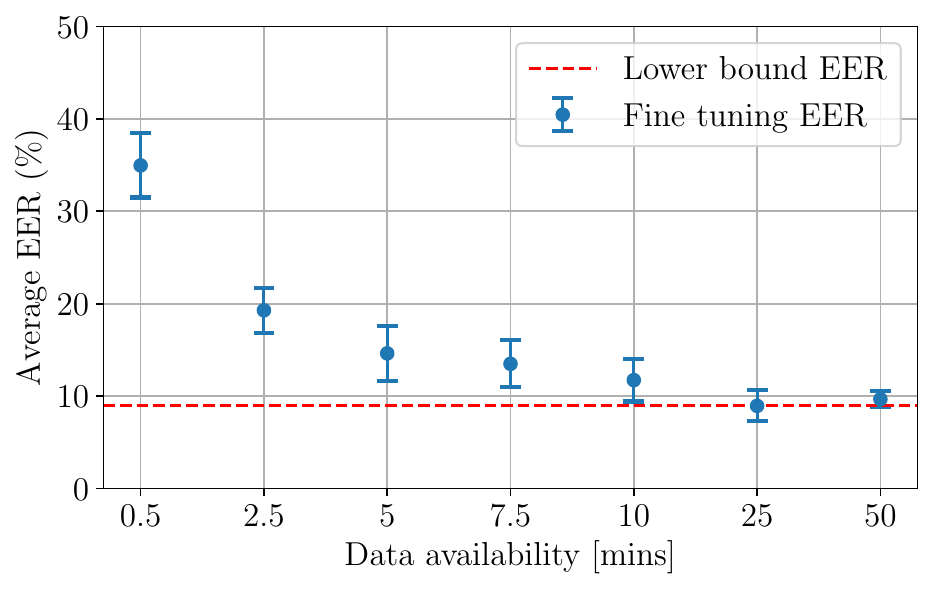}\label{fig:FT1}}
     \subfloat[Source \acp{ID} $(2, 3)$ and target \acp{ID} $(12, 15)$.]{\includegraphics[width=0.25\linewidth]{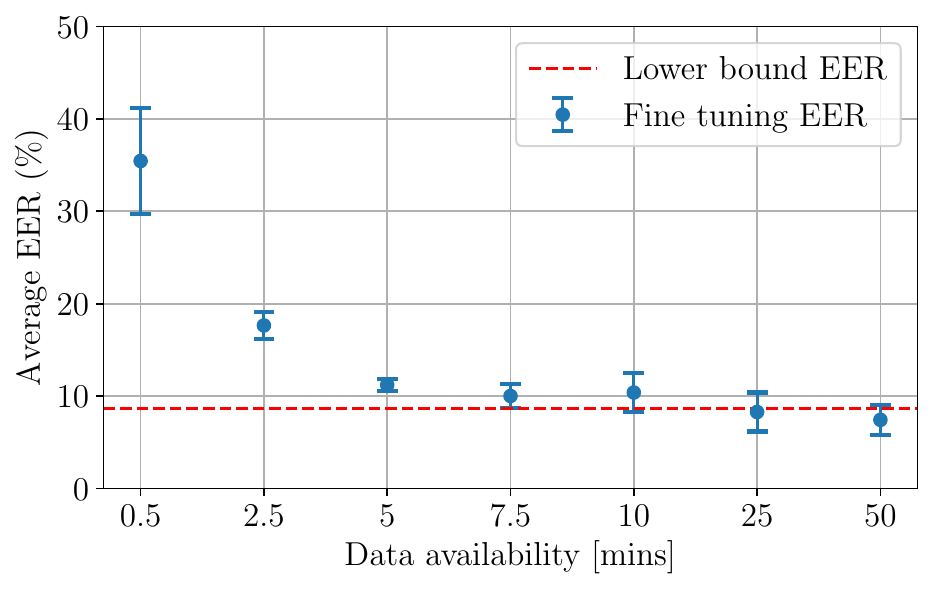}\label{fig:FT2}}
     \subfloat[Source \acp{ID} $(2, 3, 4)$ and target \acp{ID} $(12, 18, 15)$.]{\includegraphics[width=0.25\linewidth]{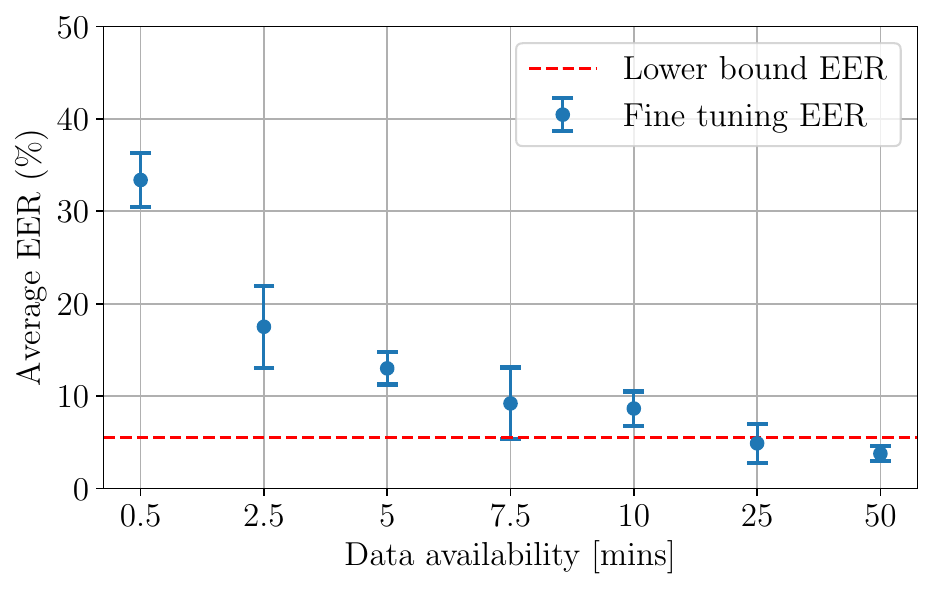}\label{fig:FT3}}
     \subfloat[Source \acp{ID} $(2, 3, 4, 5)$ and target \acp{ID} $(12, 14, 16, 17)$.]{\includegraphics[width=0.25\linewidth]{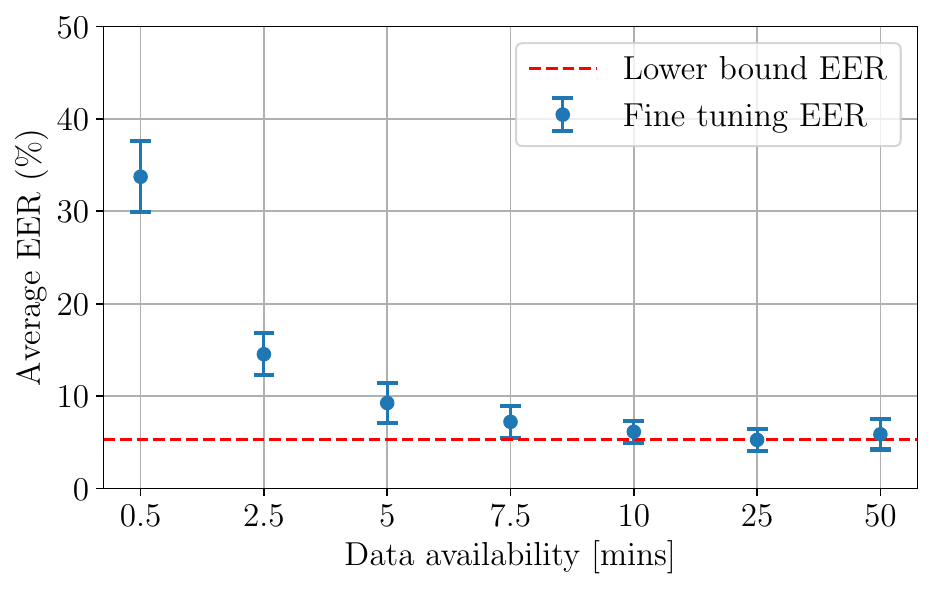}\label{fig:FT4}}
    \caption{Confidence interval of \ac{EER} of M-ALRAD by fine-tuning the model with different availabilities of training data for (a) one, (b) two, (c) three, and (d) four microphones.}
    \label{fig:fine-tuning}
\end{figure*}

\subsection{Fine-tuning with target data}
Addressing RQ2 and RQ3, we fine-tune a pretrained model on a source array $\mathrm{D}_i$ to an unseen target array $\mathrm{D}_j$, with various number of microphones. To study this adaptation, fine-tuning was performed by varying the amount of data available from target domain from half a minute to fifty minutes. Analyzing this behavior is particularly important in real-world applications where collecting large amounts of data may not always be feasible~\cite{Lin_ICASSP_2024}. In fact, from this study is possible to determine how much data is practically needed to achieve a desired level of model's performance. 

Figure~\ref{fig:fine-tuning} shows the \acp{EER} of the fine-tuned M-ALRAD for one microphone~(\ref{fig:FT1}), two microphones~(\ref{fig:FT2}), three microphones~(\ref{fig:FT3}), and four microphones~(\ref{fig:FT4}). It is worth highlighting that in all scenarios the \ac{EER} lower bound, e.g., a version of the model trained solely on target data, is reached by using at least ten minutes of target recordings. % This happens because of the reduced number of parameters of M-ALRAD (approximately $300$ K). 

\section{Conclusion} \label{sec:conc}
This study analyzed the microphone array mismatch problem in replay speech detection. Using a state-of-the-art learning-based detector, which also performs adaptive beamforming, we demonstrated that single-channel recordings suffers from generalization across different devices, while multi-channel configurations slightly improve the detection rate but remain susceptible to cross-array variability. If there is an availability of target array recordings, it is possible to adapt a pre-trained model by fine-tuning, showing that even a limited amount of target-domain data can substantially restore detection performance. Future research should explore array geometry-aware learning approaches, thus including some geometry features or a-priori knowledge of the microphone array, and domain adaptation techniques to further enhance robustness, i.e., generalization capabilities, against microphone mismatches.

\bibliographystyle{IEEEtran}
\bibliography{biblio}

\begin{thebibliography}{10}
\providecommand{\url}[1]{#1}
\csname url@samestyle\endcsname
\providecommand{\newblock}{\relax}
\providecommand{\bibinfo}[2]{#2}
\providecommand{\BIBentrySTDinterwordspacing}{\spaceskip=0pt\relax}
\providecommand{\BIBentryALTinterwordstretchfactor}{4}
\providecommand{\BIBentryALTinterwordspacing}{\spaceskip=\fontdimen2\font plus
\BIBentryALTinterwordstretchfactor\fontdimen3\font minus \fontdimen4\font\relax}
\providecommand{\BIBforeignlanguage}[2]{{%
\expandafter\ifx\csname l@#1\endcsname\relax
\typeout{** WARNING: IEEEtran.bst: No hyphenation pattern has been}%
\typeout{** loaded for the language `#1'. Using the pattern for}%
\typeout{** the default language instead.}%
\else
\language=\csname l@#1\endcsname
\fi
#2}}
\providecommand{\BIBdecl}{\relax}
\BIBdecl

\bibitem{Huang_IoTJ_2022}
W.~Huang, W.~Tang, H.~Jiang, J.~Luo, and Y.~Zhang, ``Stop deceiving! an effective defense scheme against voice impersonation attacks on smart devices,'' \emph{IEEE Internet of Things Journal}, vol.~9, no.~7, pp. 5304--5314, 2022.

\bibitem{Liu_TASLP_2023}
X.~Liu, X.~Wang, M.~Sahidullah, J.~Patino, H.~Delgado, T.~Kinnunen, M.~Todisco, J.~Yamagishi, N.~Evans, A.~Nautsch, and K.~A. Lee, ``{ASVspoof 2021: Towards Spoofed and Deepfake Speech Detection in the Wild},'' \emph{IEEE/ACM Transactions on Audio, Speech, and Language Processing}, vol.~31, pp. 2507--2522, 2023.

\bibitem{Kinnunen_ICASSP_2017}
T.~Kinnunen, M.~Sahidullah, M.~Falcone, L.~Costantini, R.~G. Hautamäki, D.~Thomsen, A.~Sarkar, Z.~Tan, H.~Delgado, M.~Todisco, N.~Evans, V.~Hautamäki, and K.~A. Lee, ``{RedDots replayed: A new replay spoofing attack corpus for text-dependent speaker verification research},'' in \emph{IEEE International Conference on Acoustics, Speech and Signal Processing (ICASSP)}, 2017.

\bibitem{Gong_SPL_2020}
Y.~Gong, J.~Yang, and C.~Poellabauer, ``{Detecting replay attacks using multi-channel audio: A neural network-based method},'' \emph{IEEE Signal Processing Letters}, vol.~27, pp. 920--924, 2020.

\bibitem{Delac_ISEM_2004}
K.~Delac and M.~Grgic, ``A survey of biometric recognition methods,'' in \emph{Proceedings. Elmar-2004. 46th International Symposium on Electronics in Marine}, 2004.

\bibitem{Neri_2025_ARXIV}
M.~Neri and T.~Virtanen, ``Multi-channel replay speech detection using an adaptive learnable beamformer,'' \emph{IEEE Open Journal of Signal Processing}, pp. 1--7, 2025.

\bibitem{Gong_Interspeech_2019}
Y.~Gong, J.~Yang, J.~Huber, M.~MacKnight, and C.~Poellabauer, ``{ReMASC: Realistic Replay Attack Corpus for Voice Controlled Systems},'' \emph{Interspeech}, 2019.

\bibitem{kinnunen2017asvspoof}
T.~Kinnunen, M.~Sahidullah, H.~Delgado, M.~Todisco, N.~Evans, J.~Yamagishi, and K.~A. Lee, ``{The ASVspoof 2017 challenge: Assessing the limits of replay spoofing attack detection},'' in \emph{Interspeech}, 2017.

\bibitem{todisco2019asvspoof}
M.~Todisco, X.~Wang, V.~Vestman, M.~Sahidullah, H.~Delgado, A.~Nautsch, J.~Yamagishi, N.~Evans, T.~Kinnunen, and K.~A. Lee, ``{ASVspoof 2019: Future horizons in spoofed and fake audio detection},'' in \emph{Interspeech}, 2019.

\bibitem{Luo_ICASSP_2021}
A.~Luo, E.~Li, Y.~Liu, X.~Kang, and Z.~J. Wang, ``{A Capsule Network Based Approach for Detection of Audio Spoofing Attacks},'' in \emph{IEEE International Conference on Acoustics, Speech and Signal Processing (ICASSP)}, 2021.

\bibitem{Xue_LSP_2024}
J.~Xue, C.~Fan, J.~Yi, J.~Zhou, and Z.~Lv, ``{Dynamic Ensemble Teacher-Student Distillation Framework for Light-Weight Fake Audio Detection},'' \emph{IEEE Signal Processing Letters}, vol.~31, pp. 2305--2309, 2024.

\bibitem{boyd2023voice}
J.~Boyd, M.~Fahim, and O.~Olukoya, ``{Voice spoofing detection for multiclass attack classification using deep learning},'' \emph{Machine Learning With Applications}, vol.~14, p. 100503, 2023.

\bibitem{Xu_TASLP_2023}
L.~Xu, J.~Yang, C.~H. You, X.~Qian, and D.~Huang, ``{Device Features Based on Linear Transformation With Parallel Training Data for Replay Speech Detection},'' \emph{IEEE/ACM Transactions on Audio, Speech, and Language Processing}, vol.~31, pp. 1574--1586, 2023.

\bibitem{heittola2020acoustic}
T.~Heittola, A.~Mesaros, and T.~Virtanen, ``Acoustic scene classification in dcase 2020 challenge: generalization across devices and low complexity solutions,'' in \emph{DCASE}, 2020.

\bibitem{Dohi2022}
K.~Dohi, K.~Imoto, N.~Harada, D.~Niizumi, Y.~Koizumi, T.~Nishida, H.~Purohit, R.~Tanabe, T.~Endo, M.~Yamamoto, and Y.~Kawaguchi, ``{Description and Discussion on DCASE 2022 Challenge Task 2: Unsupervised Anomalous Sound Detection for Machine Condition Monitoring Applying Domain Generalization Techniques},'' in \emph{DCASE}, 2022.

\bibitem{Neri_2024_ICMEW}
M.~Neri and M.~Carli, ``{Semi-Supervised Acoustic Scene Classification Under Domain Shift Using an Attention Module and Angular Loss},'' in \emph{IEEE International Conference on Multimedia and Expo Workshops (ICMEW)}, 2024.

\bibitem{omologo2001speech}
M.~Omologo, M.~Matassoni, and P.~Svaizer, ``{Speech recognition with microphone arrays},'' in \emph{Microphone arrays: signal processing techniques and applications}.\hskip 1em plus 0.5em minus 0.4em\relax Springer, 2001, pp. 331--353.

\bibitem{vincent2017analysis}
E.~Vincent, S.~Watanabe, A.~A. Nugraha, J.~Barker, and R.~Marxer, ``{An analysis of environment, microphone and data simulation mismatches in robust speech recognition},'' \emph{Computer Speech \& Language}, vol.~46, pp. 535--557, 2017.

\bibitem{Lin_ICASSP_2024}
J.~Lin, N.~Moritz, Y.~Huang, R.~Xie, M.~Sun, C.~Fuegen, and F.~Seide, ``{AGADIR: Towards Array-Geometry Agnostic Directional Speech Recognition},'' in \emph{IEEE International Conference on Acoustics, Speech and Signal Processing (ICASSP)}, 2024.

\bibitem{Han_ICASSP_2024}
C.~Han, K.~Wilson, S.~Wisdom, and J.~R. Hershey, ``{Unsupervised Multi-Channel Separation And Adaptation},'' in \emph{IEEE International Conference on Acoustics, Speech and Signal Processing (ICASSP)}, 2024.

\bibitem{Mariotte_TASLP_2024}
T.~Mariotte, A.~Larcher, S.~Montrésor, and J.~Thomas, ``{Channel-Combination Algorithms for Robust Distant Voice Activity and Overlapped Speech Detection},'' \emph{IEEE/ACM Transactions on Audio, Speech, and Language Processing}, vol.~32, pp. 1859--1872, 2024.

\bibitem{Heikkinen_ICASSP_2024}
M.~Heikkinen, A.~Politis, and T.~Virtanen, ``{Neural Ambisonics Encoding For Compact Irregular Microphone Arrays},'' in \emph{IEEE International Conference on Acoustics, Speech and Signal Processing (ICASSP)}, 2024.

\bibitem{Pan_TASLP_2025}
C.~Pan, J.~Chen, and J.~Benesty, ``An approach to microphone array geometry transform,'' \emph{IEEE Transactions on Audio, Speech and Language Processing}, pp. 1--14, 2025.

\bibitem{Neri_TASLP_2024}
M.~Neri, A.~Politis, D.~A. Krause, M.~Carli, and T.~Virtanen, ``{Speaker Distance Estimation in Enclosures From Single-Channel Audio},'' \emph{IEEE/ACM Transactions on Audio, Speech, and Language Processing}, vol.~32, pp. 2242--2254, 2024.

\bibitem{Neri_WASPAA_2023}
M.~Neri, A.~Politis, D.~A. Krause, M.~Carli, and T.~Virtanen, ``{Single-Channel Speaker Distance Estimation in Reverberant Environments},'' in \emph{IEEE Workshop on Applications of Signal Processing to Audio and Acoustics (WASPAA)}, 2023.

\end{thebibliography}

\end{document}